\documentclass[aps,prc,imp,preprint,superscriptaddress]{revtex4}
\usepackage[pdftex,
 letterpaper=true,
 hyperindex=true,
 breaklinks=true,
 colorlinks=false,
 citecolor=blue,
 pdftitle={},
 pdfauthor={}]
{hyperref}
\usepackage{graphicx}
\begin{document}

\title{Measurement of the $^{90, 92}$Zr(p,$\gamma$)$^{91,93}$Nb reactions for the nucleosynthesis of elements around A=90}

\author{A.~Spyrou}
    \email[]{spyrou@nscl.msu.edu}
	\affiliation{National Superconducting Cyclotron Laboratory, Michigan State University, East Lansing, Michigan 48824, USA}
	\affiliation{Department of Physics \& Astronomy, Michigan State University, East Lansing, Michigan 48824, USA}
	\affiliation{Joint Institute for Nuclear Astrophysics, Michigan State University, East Lansing, MI 48824, USA}

\author{S.~J.~Quinn}
	\affiliation{National Superconducting Cyclotron Laboratory, Michigan State University, East Lansing, Michigan 48824, USA}
	\affiliation{Department of Physics \& Astronomy, Michigan State University, East Lansing, Michigan 48824, USA}
	\affiliation{Joint Institute for Nuclear Astrophysics, Michigan State University, East Lansing, MI 48824, USA}

\author{A.~Simon}
	\affiliation{National Superconducting Cyclotron Laboratory, Michigan State University, East Lansing, Michigan 48824, USA}
	\affiliation{Joint Institute for Nuclear Astrophysics, Michigan State University, East Lansing, MI 48824, USA}

\author{T.~Rauscher}
	\affiliation{Centre for Astrophysical Research, School of Physics, Astronomy, and Mathematics, University of Hertfordshire, Hatfield AL10 9AB, United Kingdom}
	\affiliation{Institute of Nuclear Research (ATOMKI), H-4001 Debrecen, POB 51, Hungary}
	\affiliation{Department of Physics, University of Basel, 4056 Basel, Switzerland}

\author{A.~Battaglia}
	\affiliation{Department of Physics and The Joint Institute for Nuclear Astrophysics, University of Notre Dame, Notre Dame, Indiana 46556, USA}

\author{A.~Best}
	\affiliation{Department of Physics and The Joint Institute for Nuclear Astrophysics, University of Notre Dame, Notre Dame, Indiana 46556, USA}
	\altaffiliation[Present Address: ]{Lawrence Berkeley National Laboratory, Berkeley, CA 94720, USA}

\author{B.~Bucher}
	\affiliation{Department of Physics and The Joint Institute for Nuclear Astrophysics, University of Notre Dame, Notre Dame, Indiana 46556, USA}

\author{M.~Couder}
	\affiliation{Department of Physics and The Joint Institute for Nuclear Astrophysics, University of Notre Dame, Notre Dame, Indiana 46556, USA}

\author{P.~A.~DeYoung}
	\affiliation{Department of Physics, Hope College, Holland, Michigan 49423, USA}

\author{A.~C.~Dombos}
	\affiliation{National Superconducting Cyclotron Laboratory, Michigan State University, East Lansing, Michigan 48824, USA}
	\affiliation{Department of Physics \& Astronomy, Michigan State University, East Lansing, Michigan 48824, USA}
	\affiliation{Joint Institute for Nuclear Astrophysics, Michigan State University, East Lansing, MI 48824, USA}

\author{X.~Fang}
	\affiliation{Department of Physics and The Joint Institute for Nuclear Astrophysics, University of Notre Dame, Notre Dame, Indiana 46556, USA}

\author{J.~G\"orres}
	\affiliation{Department of Physics and The Joint Institute for Nuclear Astrophysics, University of Notre Dame, Notre Dame, Indiana 46556, USA}

\author{A.~Kontos}
	\affiliation{Department of Physics and The Joint Institute for Nuclear Astrophysics, University of Notre Dame, Notre Dame, Indiana 46556, USA}

\author{Q.~Li}
	\affiliation{Department of Physics and The Joint Institute for Nuclear Astrophysics, University of Notre Dame, Notre Dame, Indiana 46556, USA}

\author{L.~Y.~Lin}
	\affiliation{National Superconducting Cyclotron Laboratory, Michigan State University, East Lansing, Michigan 48824, USA}
	\affiliation{Department of Physics \& Astronomy, Michigan State University, East Lansing, Michigan 48824, USA}

\author{A.~Long}
	\affiliation{Department of Physics and The Joint Institute for Nuclear Astrophysics, University of Notre Dame, Notre Dame, Indiana 46556, USA}

\author{S.~Lyons}
	\affiliation{Department of Physics and The Joint Institute for Nuclear Astrophysics, University of Notre Dame, Notre Dame, Indiana 46556, USA}

\author{B.~S.~Meyer}
	\affiliation{Department of Physics and Astronomy, Clemson University, 118 Kinard Laboratory, Clemson, SC 29634-0978, USA}

\author{A.~Roberts}
	\affiliation{Department of Physics and The Joint Institute for Nuclear Astrophysics, University of Notre Dame, Notre Dame, Indiana 46556, USA}

\author{D.~Robertson}
	\affiliation{Department of Physics and The Joint Institute for Nuclear Astrophysics, University of Notre Dame, Notre Dame, Indiana 46556, USA}

\author{K.~Smith}
	\affiliation{Department of Physics and The Joint Institute for Nuclear Astrophysics, University of Notre Dame, Notre Dame, Indiana 46556, USA}

\author{M.~K.~Smith}
	\affiliation{Department of Physics and The Joint Institute for Nuclear Astrophysics, University of Notre Dame, Notre Dame, Indiana 46556, USA}

\author{E.~Stech}
	\affiliation{Department of Physics and The Joint Institute for Nuclear Astrophysics, University of Notre Dame, Notre Dame, Indiana 46556, USA}

\author{B. Stefanek}
	\affiliation{National Superconducting Cyclotron Laboratory, Michigan State University, East Lansing, Michigan 48824, USA}
	\affiliation{Department of Physics \& Astronomy, Michigan State University, East Lansing, Michigan 48824, USA}

\author{W.~P.~Tan}
	\affiliation{Department of Physics and The Joint Institute for Nuclear Astrophysics, University of Notre Dame, Notre Dame, Indiana 46556, USA}

\author{X.~D.~Tang}
	\affiliation{Department of Physics and The Joint Institute for Nuclear Astrophysics, University of Notre Dame, Notre Dame, Indiana 46556, USA}

\author{M.~Wiescher}
	\affiliation{Department of Physics and The Joint Institute for Nuclear Astrophysics, University of Notre Dame, Notre Dame, Indiana 46556, USA}

\date{\today}

\begin{abstract}

Cross section measurements of the reactions $^{90, 92}$Zr(p,$\gamma$)$^{91,93}$Nb were 
performed using the NSCL SuN detector at the University of Notre Dame. These reactions are part of the nuclear
reaction flow for the synthesis of the light p nuclei. For the $^{90}$Zr(p,$\gamma$)$^{91}$Nb reaction 
the new measurement resolves the disagreement between previous results. For the $^{92}$Zr(p,$\gamma$)$^{93}$Nb reaction
the present work reports the first measurement of this reaction cross section. Both reaction cross sections are 
compared to theoretical calculations and a very good agreement with the standard NON-SMOKER model is observed.

\end{abstract}

\maketitle


\section{Introduction}
\label{sec:INTRO}

The origin of the heavy elements remains one of the 
overarching  questions for the nuclear astrophysics community.
The general concepts of heavy element nucleosynthesis were introduced in
1957 by the famous B$^2$FH publication \cite{Bur57}, however, 55 years later, 
many of the details of the responsible astrophysical processes are still not well
understood.

A small fraction of the heavy elements is located on the neutron deficient 
side of the valley of stability and forms the group of the so-called p nuclei \cite{Arn03, Rau13}.
This group consists of 35 stable isotopes in the mass region between Se (Z=34)
and Hg (Z=80) that cannot be synthesized by the two neutron-capture-induced processes
(s- and r-process) (e.g. \cite{Mey94}). Their synthesis requires a different astrophysical mechanism
traditionally called the ``p process''. It is not yet clear whether the p process is a single 
or multiple independent astrophysical scenarios. A number of astrophysical settings
have been proposed for the site of the production of the p nuclei, such as type II supernovae 
(SNII) \cite{Ray90, Rau02}, supernovae type Ia (e.g. \cite{Tra11} and references therein) , 
the $\nu$p process in the neutrino driven winds  of SNII \cite{Fro06, Wan11}, the rp-process in matter accreted on
the surface of neutron stars \cite{Sch98} and others \cite{Arn03, Rau13}. 

The most favored scenario, and the one that has been investigated the most to date, 
takes place in SNII when the shock front passes through the O/Ne-rich
layers of the massive star \cite{Ray90, Rau02}. In this scenario, the main reactions responsible for the formation of the 
p nuclei are neutron, proton and alpha-particle photodisintegrations together 
with a possible contribution to the light p nuclei from proton-captures at the higher temperatures.
Due to the dominance of photodisintegration reactions, this process is also called ``$\gamma$ process'' and
it takes place at temperatures between 1.5 and 3.5 GK.  This scenario, although able to produce
p nuclei in the whole mass range, is not able to reproduce the abundance patterns 
observed in solar system samples. Large underproductions relative to the solar abundance pattern occur in the lighter masses, especially for 
$^{92,94}$Mo and $^{96,98}$Ru. Many authors investigated 
the possible sources of uncertainties in the predicted abundance distributions (\cite{Arn03, Rau13} and references therein)
and it was shown that these come both from uncertainties in the stellar conditions as well as the nuclear physics input.
The present work focuses on the nuclear physics aspect of p-process nucleosynthesis via the investigation of 
the $^{90,92}$Zr(p,$\gamma$)$^{91,93}$Nb reactions, in the region around the puzzling $^{92,94}$Mo and $^{96,98}$Ru.
 
Astrophysical calculations were performed using the post-processing code available in NucNet tools \cite{NucNet}, a suite of nuclear reaction codes developed at Clemson University. 
In the calculations the $\gamma$ process was investigated using a full nuclear reaction network for a type II supernova explosion when the shock front passes
through the O/Ne layer of a 25~M$_\odot$ star. The calculations were performed in a multi-layer model (11 layers) using the seed distribution of a pre-explosion 25 M$_\odot$ star. 
The seed distribution and temperature and density profiles were taken as described in Ref.~\cite{Rap06}. In this model the calculations show that the p nuclei are produced mainly through photodisintegration 
reactions, as expected; however, for the lighter p nuclei there is an important contribution coming 
from (p,$\gamma$) reactions that take place in the inner-most layers, i.e. highest temperature/density regions of the O/Ne layer. This contribution was already observed in the sensitivity study
by Rapp {\it et al.} \cite{Rap06} and was explored in recent studies of the $^{74}$Ge(p,$\gamma$)$^{75}$As reaction \cite{Sau12, Qui13}. Similarly, the two reactions studied in the present work 
could have a direct contribution to the production of $^{92,94}$Mo via two consecutive (p,$\gamma$) reactions from $^{90,92}$Zr. These (p,$\gamma$) reactions could have a contribution at the 
highest temperatures, especially if the reaction rates are significantly higher than the ones currently used in astrophysical calculations presented in the JINA REACLIB database \cite{REACLIB}.

In the literature there are two previous measurements of the $^{90}$Zr(p,$\gamma$)$^{91}$Nb reaction. One was a thick-target measurement 
 using the activation technique by Roughton {\it et al.} \cite{Rou79}. The authors were able to extract an astrophysical reaction rate from their thick-target measurements
 without the need to first calculate the reaction cross section, which, as mentioned in their paper, would introduce additional uncertainties. 
The second measurement was performed in-beam by Laird {\it et al.} \cite{Lai87}  by measuring the $\gamma$ transitions populating the ground state, 
or the first excited metastable state of the final nucleus $^{91}$Nb. The individual contributions of the $\gamma$ transitions were summed to 
give the total (p,$\gamma$) cross section. Since the two works did not produce the same final quantity it is hard to compare them directly. Nevertheless, comparison to the
same theoretical calculations (section \ref{sec:DISC}) show that the two measurements are not in agreement. 
For this reason, in the present work an independent measurement of the $^{90}$Zr(p,$\gamma$)$^{91}$Nb reaction cross section
was performed.  The results of the two previous measurements and of the present one will be discussed in section \ref{sec:DISC}. 
The $^{92}$Zr(p,$\gamma$)$^{93}$Nb reaction has not been measured before.

The present paper is organized as follows: the experimental setup and the technique are presented in section \ref{sec:EXP}, the results for the two reactions
of interest are shown in section \ref{sec:RESULTS}, the discussion of these results, comparison to theoretical calculations and astrophysical implications are presented in 
section \ref{sec:DISC} and the paper is completed with the conclusions in section \ref{sec:CONC}.


\section{Experimental}
\label{sec:EXP}

The experiment was performed at the FN Tandem Accelerator of the University of Notre Dame. 
A proton beam at energies between 2.0 and 5.0 MeV (uncertainty $\approx$2~keV) interacted with the two self-supporting and isotopically 
enriched Zr targets. The target thickness was measured using the Rutherford Back-Scattering (RBS)
technique at the Hope College Ion Beam Analysis Laboratory (HIBAL). 
For the RBS measurements, a 2.95~MeV alpha beam was used. After impinging on the target the scattered alpha particles were
detected at 168.2$^\circ$ in a Si surface barrier detector with a 40~keV energy resolution. The RBS spectra were simulated using the software SIMNRA \cite{SIMNRA}, 
where all the parameters of the experiment were fixed and the only free parameter was the target thickness. 
The measured thickness
was 969(48)~$\mu$g/cm$^2$ for the $^{90}$Zr target and 960(47)~$\mu$g/cm$^2$ for the $^{92}$Zr one. The two
targets were highly enriched in the isotope of interest (98(1)\%).
The beam current on target was at most 22~nA to minimize possible dead-time effects. The total accumulated
charge was between 1 and 10~$\mu$C.

The cross section measurements were performed using the newly commissioned Summing NaI(Tl) (SuN) detector
of the National Superconducting Cyclotron Laboratory, Michigan State University. SuN is a barrel shaped NaI(Tl)
scintillator, 16~inches in diameter and 16~inches in height with a bore hole along its axis. A detailed description of 
SuN, the electronics setup, and data acquisition system can be found in Ref.~\cite{Sim13}. 

The targets were mounted in the center
of the detector where the large angular coverage and high detection efficiency of SuN allowed for the use of the $\gamma$-summing 
technique \cite{Spy07, Sim13}. In this technique the capture of a proton by the target nucleus is measured by summing
the $\gamma$ rays emitted during the de-excitation of the produced nucleus. The energy of the entry state
of the produced nucleus is $E_{\Sigma} = Q + E_{cm}$, where $Q$ is the Q-value of the reaction  and $E_{cm}$ is the center-of-mass energy of the initial system.
In the summing technique the $\gamma$ spectrum is expected to show the so-called ``sum peak'', located at the energy $E_{\Sigma}$, 
in the high energy region.

\begin{figure}[t]
\includegraphics[width=0.5\textwidth]{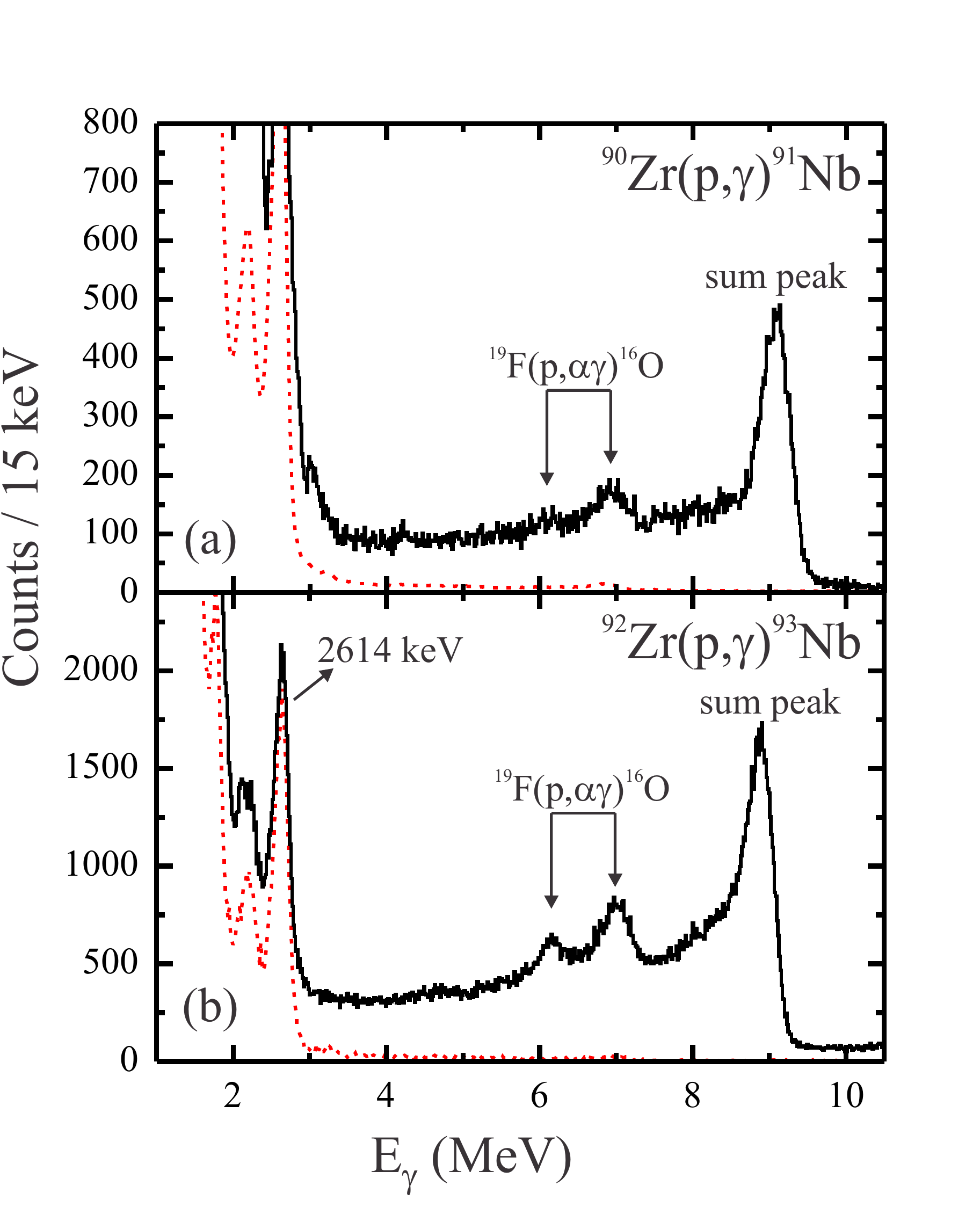}
\caption{\label{fig:spectra} (Color online) Spectra taken with the SuN detector for (a) the $^{90}$Zr(p,$\gamma$)$^{91}$Nb 
reaction at proton beam energy of 4~MeV and (b) the $^{92}$Zr(p,$\gamma$)$^{93}$Nb reaction at 2.8~MeV. In both figures the black-solid
line represents the beam-on-target spectrum while the red-dashed line shows the normalized room background. }
\end{figure}

Typical spectra for the $^{90}$Zr(p,$\gamma$)$^{91}$Nb and $^{92}$Zr(p,$\gamma$)$^{93}$Nb
reactions are shown in Fig. \ref{fig:spectra}. The high energy part of the spectra is dominated by the sum peak for 
each of the reactions. For the case of the $^{90}$Zr(p,$\gamma$)$^{91}$Nb 
reaction shown in Fig.~\ref{fig:spectra}a, the Q-value is 5.16~MeV and the beam energy was 4~MeV resulting in a sum peak at 9.09~MeV (assuming that the interaction takes place at the center of the target). 
For the case of the $^{92}$Zr(p,$\gamma$)$^{93}$Nb 
reaction shown in Fig.~\ref{fig:spectra}b, the Q-value is 6.04~MeV and the beam energy was 2.8~MeV resulting in a sum peak at 8.78~MeV. 
 In the region between 6 and 7~MeV the spectrum is contaminated by the $\gamma$ rays
from the reaction $^{19}$F(p,$\alpha \gamma$)$^{16}$O, which is a common contaminant reaction in proton induced
measurements. The low energy region of the spectrum, below 3~MeV, is dominated by the room background $\gamma$
rays from $^{40}$K at 1461~keV and from the Th decay chain at 2614~keV. 
 The presence of any beam-induced background will also contribute to
the low-energy regions of the spectra. The only source of background
in the high-energy region, where the sum peak is observed, comes from
cosmic rays. For all measurements presented in 
this work the cosmic ray background contribution was negligible.


\section{Results}
\label{sec:RESULTS}

The reaction cross section can be determined from the analysis of the sum peak in each of the spectra. The peak integration was based on
 3$\sigma$ limits on each side of the centroid of the sum peak and linear background subtraction. This method was chosen to be consistent with the 
 efficiency calibration procedure described in detail in Ref.~\cite{Sim13}. In both reactions presented in this work the final 
 nucleus also has a long-lived metastable state, which will not be summed in the main sum peak. In this case, a second sum peak is expected in the spectra
 that will be located at an energy equal to the difference between the entry state and the metastable state \cite{Spy07, Har13}. For the $^{90}$Zr(p,$\gamma$)$^{91}$Nb
 reaction, the metastable state in $^{91}$Nb is located at 104.6~keV with a half-life of 60.9~d \cite{NNDC}. For the $^{92}$Zr(p,$\gamma$)$^{93}$Nb  
 reaction, the metastable state in $^{93}$Nb is located at 30.8~keV with a half-life of 16.1~y \cite{NNDC}. In both cases
 the metastable-state sum peak is expected to be overlapping with the ground-state sum peak and show a small extension on the low energy side, which 
 was taken into account in the analysis. 
 
 Using the summing technique, the reaction cross section can be calculated from the analyzed spectra with:
 \begin{equation}
 \sigma = \frac{A}{N_A N_b}  \frac{1}{\xi} \frac{I_\Sigma}{\epsilon_\Sigma}
 \end{equation}
where A is the atomic weight of the target nucleus (in amu), $N_A$ is the Avogadro constant, $N_b$ is the total number 
of beam particles impinging on the target, $\xi$ is the target thickness, $I_\Sigma$ is the sum-peak intensity and $\epsilon_\Sigma$ 
is the sum-peak efficiency. $N_b$ was measured during the experiment using a current integrator, $\xi$ was determined
as mentioned before by means of RBS analysis, and $I_\Sigma$ was the result of the sum-peak analysis described in the previous paragraph. 
The sum peak efficiency $\epsilon_\Sigma$  depends on the 
sum-peak energy but also on the multiplicity of the $\gamma$ cascade. Two techniques have been developed in the literature for 
the experimental determination of $\epsilon_\Sigma$, the ``In/Out ratio'' method \cite{Spy07} and the ``hit pattern'' technique \cite{Sim13}.
In the present work the sum-peak efficiency was determined using the latter method, which was developed and optimized specifically for the SuN detector. 
In this method, the segmentation of SuN provides a sensitive tool for determining the $\gamma$-cascade multiplicity. The number of SuN segments
that recorded a signal in each event depends on the number of emitted $\gamma$ rays and on the sum-peak energy. Details on the ``hit pattern" technique
were given in Ref.~\cite{Sim13}. The efficiencies for the two reactions of interest ranged from 21.0(2.1)\% to 38.5(3.0)\%. 

The cross section results for the two reactions of interest are presented in Tables~\ref{tab:Zr90cs} and \ref{tab:Zr92cs}.
In both tables the first column presents the proton beam energy. The second column shows the center-of-mass 
energy with the assumption that the reaction is taking place at the center of the target. The energy uncertainty due to this
assumption and due to a 2~keV uncertainty in the set beam energy is also presented in the second column. 
The third column presents the reaction cross section and the last column 
the astrophysical S-factor. The latter quantity was calculated as:

 \begin{equation}
 S = E \sigma(E) e^{2\pi\eta} 
 \end{equation}

where E is the center-of-mass energy, $\sigma(E)$ is the reaction cross section and $\eta= Z_1Z_2e^2/\hbar v$ is the Sommerfeld parameter, with $Z_1$ and $Z_2$ the proton
numbers of the interacting particles and $v$ their relative velocity. The uncertainties presented in the two tables include: 5\% for the target thickness estimate,
5\% for the charge accumulation, of the order of 1\% statistical uncertainty and approximately 10\% uncertainty based on the efficiency analysis in \cite{Sim13}. 

\begin{table}[h]
\centering
\caption{Cross sections and S-factors of the present work for the reaction $^{90}$Zr(p,$\gamma$)$^{91}$Nb.}
\begin{tabular*}{0.45\textwidth}{c@{\extracolsep{\fill}}ccc}
\hline \hline
$E_{beam}$ & $E_{c.m.}$ & $\sigma$ & S-factor\\
(keV) & (keV) & (mb) & ($10^6$ keV barn)\\
\hline
5000	&	4925 $\pm$ 20	&	6.66 $\pm$ 0.99   &   1.65 $\pm$ 0.25  \\
4800	&	4727 $\pm$ 20	&	5.77 $\pm$ 0.69   &   1.98 $\pm$ 0.24  \\
4600	&	4529 $\pm$ 20	&	4.42 $\pm$ 0.60   &   2.15 $\pm$ 0.29  \\
4400	&	4331 $\pm$ 21	&	3.65 $\pm$ 0.47   &   2.59 $\pm$ 0.33  \\
4200	&	4132 $\pm$ 22	&	2.84 $\pm$ 0.39   &   3.01 $\pm$ 0.41  \\
4000	&	3933 $\pm$ 23	&	1.80 $\pm$ 0.29   &   2.95 $\pm$ 0.47  \\
3800	&	3735 $\pm$ 24	&	1.43 $\pm$ 0.18   &   3.73 $\pm$ 0.48  \\
3600	&	3536 $\pm$ 24	&	0.97 $\pm$ 0.14   &   4.23 $\pm$ 0.60  \\
3400	&	3338 $\pm$ 25	&	0.66 $\pm$ 0.08   &   5.00 $\pm$ 0.60  \\
3200	&	3139 $\pm$ 26	&	0.39 $\pm$ 0.05   &   5.39 $\pm$ 0.68  \\
3000	&	2940 $\pm$ 27	&	0.28 $\pm$ 0.04   &   7.54 $\pm$ 1.08  \\
2800	&	2741 $\pm$ 28	&	0.17 $\pm$ 0.02   &   9.64 $\pm$ 1.19  \\
\hline \hline
\label{tab:Zr90cs}
\end{tabular*}
\end{table}

\begin{table}[h]
\centering
\caption{Cross sections and S-factors of the present work for the reaction $^{92}$Zr(p,$\gamma$)$^{93}$Nb.}
\begin{tabular*}{0.45\textwidth}{c@{\extracolsep{\fill}}ccc}
\hline \hline
$E_{beam}$ & $E_{c.m.}$ & $\sigma$ & S-factor\\
(keV) & (keV) & (mb) & ($10^6$ keV barn)\\
\hline
5000	&	4925 $\pm$ 22	&	0.209	 $\pm$ 	0.029	  &  	0.052	 $\pm$ 	0.007	\\
4800	&	4726 $\pm$ 23	&	0.146	 $\pm$ 	0.023	  &  	0.050	 $\pm$ 	0.008	\\
4600	&	4527 $\pm$ 24	&	0.140	 $\pm$ 	0.025	  &  	0.069	 $\pm$ 	0.012	\\
4400	&	4329 $\pm$ 24	&	0.139	 $\pm$ 	0.021	  &  	0.099	 $\pm$ 	0.015	\\
4200	&	4130 $\pm$25 	&	0.117	 $\pm$ 	0.019	  &  	0.13	 $\pm$ 	0.02	\\
4000	&	3931 $\pm$ 26	&	0.094	 $\pm$ 	0.016	  &  	0.15	 $\pm$ 	0.03	\\
3800	&	3733 $\pm$ 27	&	0.107	 $\pm$ 	0.015	  &  	0.28	 $\pm$ 	0.04	\\
3600	&	3534 $\pm$ 28	&	0.099	 $\pm$ 	0.015	  &  	0.44	 $\pm$ 	0.07	\\
3400	&	3335 $\pm$ 29	&	0.072	 $\pm$ 	0.010	  &  	0.55	 $\pm$ 	0.08	\\
3200	&	3136 $\pm$ 30	&	0.115	 $\pm$ 	0.018	  &  	1.62	 $\pm$ 	0.25	\\
3000	&	2937 $\pm$ 31	&	0.484	 $\pm$ 	0.069	  &  	13.41	 $\pm$ 	1.91	\\
2800	&	2738 $\pm$ 32	&	0.405	 $\pm$ 	0.075	  &  	23.82	 $\pm$ 	4.40	\\
2600	&	2538 $\pm$ 34	&	0.215	 $\pm$ 	0.035	  &  	29.31	 $\pm$ 	4.83	\\
2400	&	2339 $\pm$ 36	&	0.093	 $\pm$ 	0.012	  &  	32.70	 $\pm$ 	4.22	\\
2200	&	2139 $\pm$ 38	&	0.036	 $\pm$ 	0.005	  &  	37.68	 $\pm$ 	5.38	\\
2000	&	1939 $\pm$ 40	&	0.016	 $\pm$ 	0.002	  &  	57.22	 $\pm$ 	7.52	\\
\hline \hline
\label{tab:Zr92cs}
\end{tabular*}
\end{table}


\section{Discussion}
\label{sec:DISC}

The $\gamma$ process temperatures quoted in section \ref{sec:INTRO}
translate to astrophysically relevant energy ranges of $1.2-4$~MeV for
$^{90}$Zr(p,$\gamma$)$^{91}$Nb and $1.2-3$ MeV (i.e., below the neutron emission
threshold) for $^{92}$Zr(p,$\gamma$)$^{93}$Nb \cite{Rau10}. The data
extend well into these ranges, allowing a detailed comparison to cross
section predictions made for the calculation of astrophysical reaction
rates. 

\begin{figure}[h]
\includegraphics[width=0.5\textwidth]{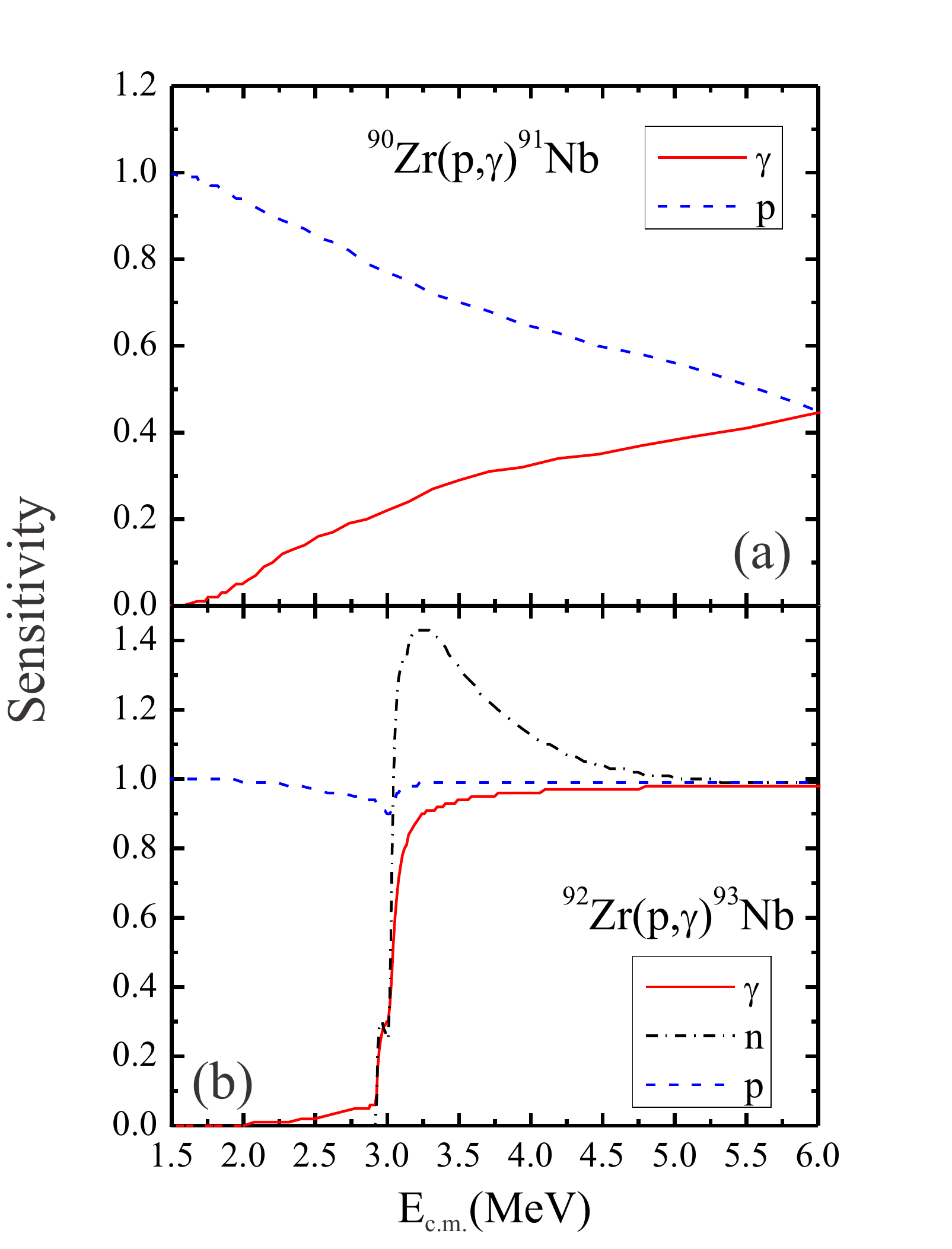}
\caption{\label{fig:sens} (Color Online) Sensitivity of the reaction $^{90}$Zr(p,$\gamma$)$^{91}$Nb (a) and 
$^{92}$Zr(p,$\gamma$)$^{93}$Nb (b) to the various widths in the Hauser-Feshbach model. 
The solid (red) line corresponds to the sensitivity to the $\gamma$ width, the dashed (blue) line shows the sensitivity to the
proton width and the dot-dashed (black) line shows the sensitivity to the neutron width. }
\end{figure}

The sensitivities of the cross sections and astrophysical $S$ factors to
variations of the averaged widths in the Hauser-Feshbach model are shown
in Fig.~\ref{fig:sens}a and \ref{fig:sens}b for the $^{90}$Zr(p,$\gamma$)$^{91}$Nb and 
$^{92}$Zr(p,$\gamma$)$^{93}$Nb reactions, respectively. 
The sensitivity $\Omega_{S_q}$ represents the impact of the variation of a model quantity $q$ (e.g. nuclear level density or averaged widths)
on the final result $\Omega$ (e.g. reaction rate, cross section or $S$-factor). The sensitivity is defined as:
\begin{equation}
\Omega_{S_q} = \frac{\upsilon_\Omega - 1}{\upsilon_q - 1} 
\end{equation} 
where $\upsilon_\Omega=\Omega_{new}/\Omega_{old}$ is the change in $\Omega$ when the quantity $q$ changes by a 
factor $\upsilon_q=q_{new}/q_{old}$. The sensitivity $\Omega_{S_q}=0$ when no change occurs and $\Omega_{S_q}=1$
when $\Omega$ changes by the same factor as $q$. More details about the sensitivity $\Omega_{S_q}$ can be found in \cite{Rau12}.

In figures \ref{fig:sens}a and \ref{fig:sens}b it can clearly be seen that the proton width is determining the cross
sections, and thus the reaction rates, at astrophysical energies.
Above the astrophysical energy window, the cross sections become
increasingly sensitive to the $\gamma$ width and also to the neutron width 
once the energy is above the neutron-emission threshold. The neutron threshold for the  $^{90}$Zr(p,$\gamma$)$^{91}$Nb
reaction is at 6.8~MeV and therefore the sensitivity to the neutron width is not observed in the energy region of Fig.~\ref{fig:sens}a.
For the  $^{92}$Zr(p,$\gamma$)$^{93}$Nb reaction, the neutron channel opens at 2.8~MeV with a clear increase in the neutron
width sensitivity at that energy as shown in Fig.~\ref{fig:sens}b.

\begin{figure}[t]
\includegraphics[width=0.5\textwidth]{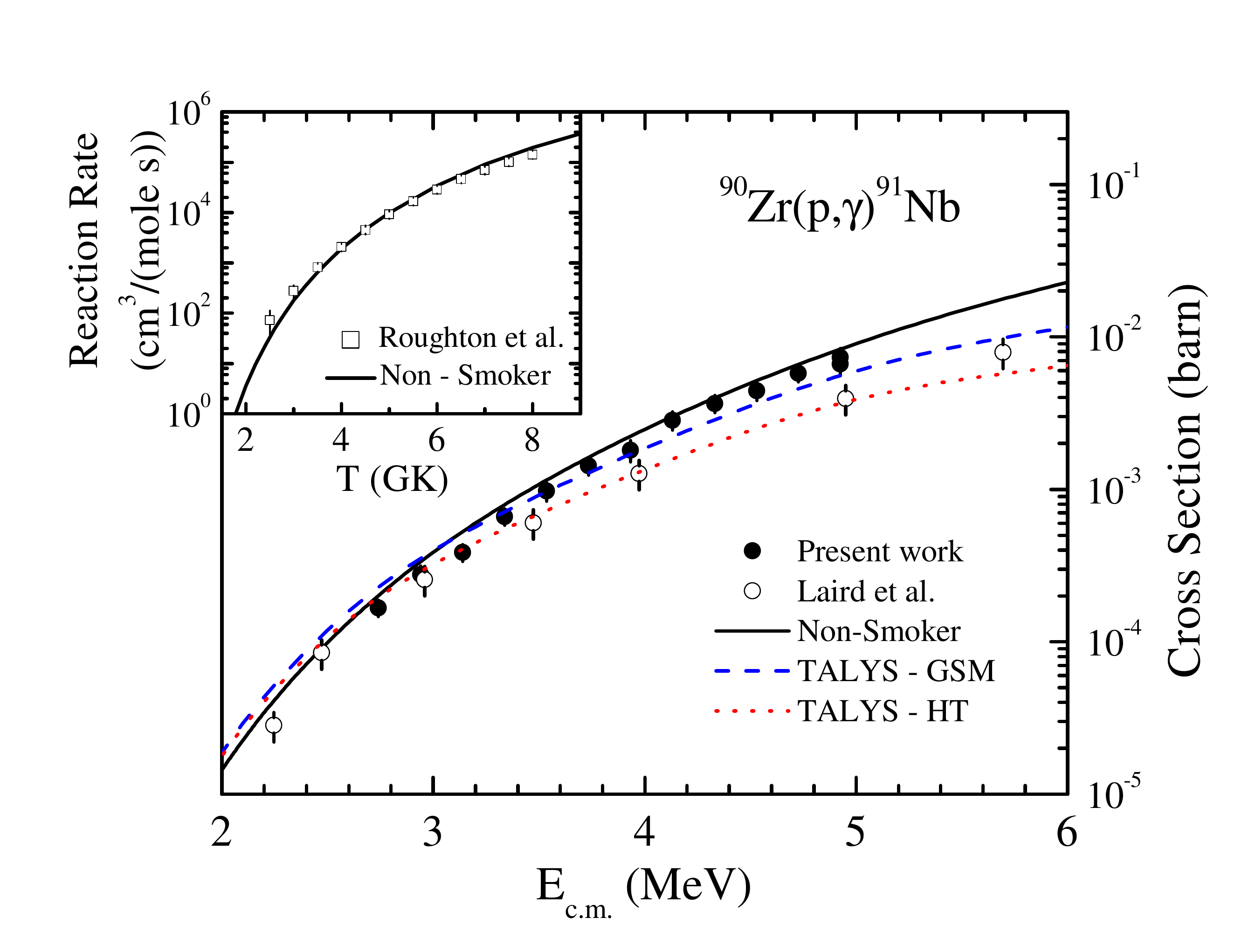}
\caption{\label{fig:Zr90cs} (Color Online) Main figure: cross section of the reaction $^{90}$Zr(p,$\gamma$)$^{91}$Nb from the present work (black dots)
and from Laird {\it et al.} \cite{Lai87} (open circles). Standard NON-SMOKER theoretical calculations \cite{Rau01} are shown in the solid (black) line and two 
TALYS  calculations \cite{Kon04} using different Nuclear Level Densities (NLD) are shown in the dashed (blue) and dotted (red) lines. See text for details. 
The inset shows a comparison of the reaction rate extracted by Roughton {\it et al.} \cite{Rou79} (open squares) compared to the standard NON-SMOKER calculation.    }
\end{figure}

\begin{figure}[b]
\includegraphics[width=0.5\textwidth]{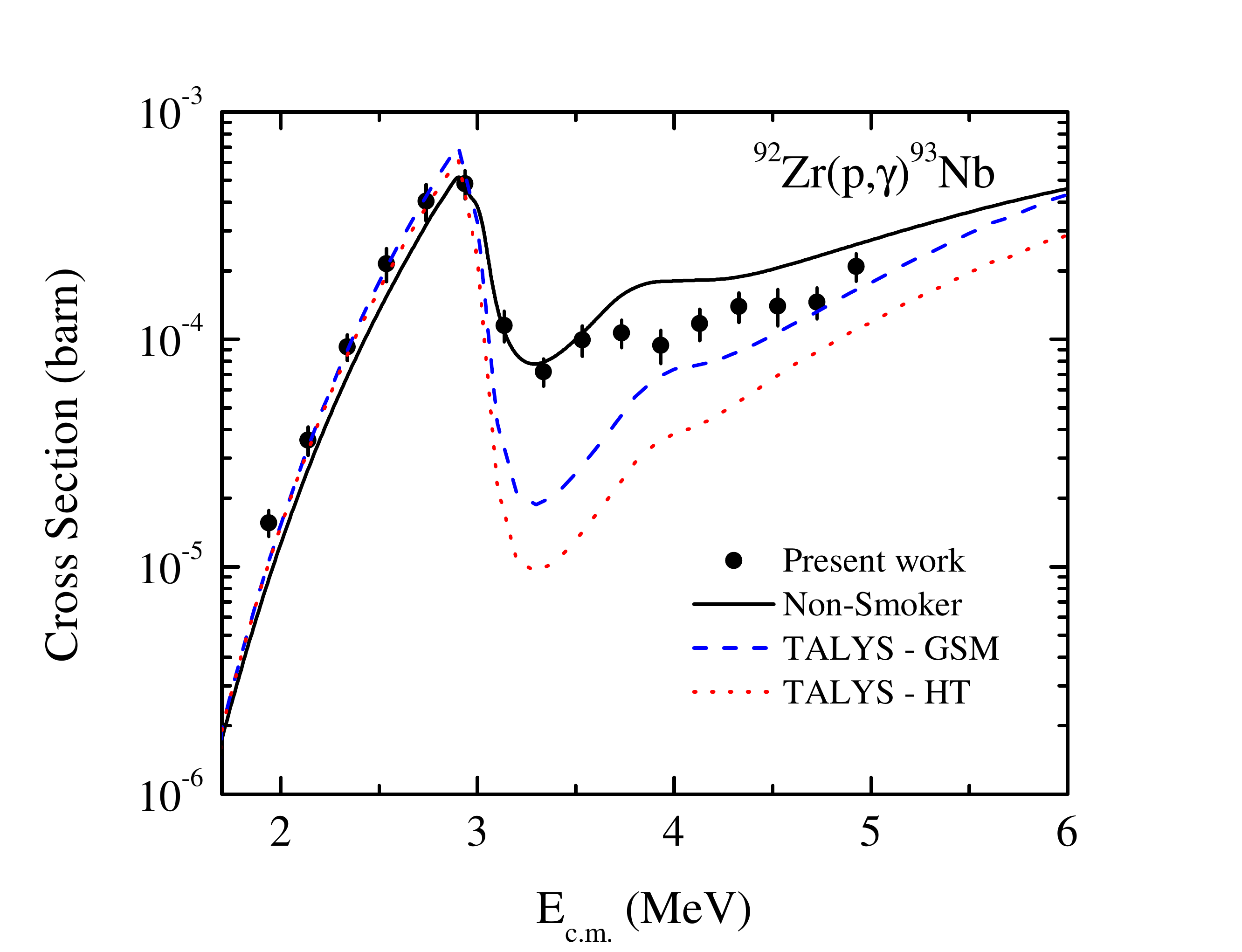}
\caption{\label{fig:Zr92cs} (Color Online) Cross section of the reaction $^{92}$Zr(p,$\gamma$)$^{93}$Nb from the present work (black dots)
compared to three theoretical calculations. Standard NON-SMOKER calculations \cite{Rau01} are shown in the solid (black) line and two 
TALYS  calculations \cite{Kon04} using different Nuclear Level Densities (NLD) are shown in the dashed (blue) and dotted (red) lines. See text for details.}
\end{figure}

The cross section measurements of the present work are compared to theoretical calculations using the codes NON-SMOKER \cite{Rau01} and 
TALYS \cite{Kon04} in Figs.~\ref{fig:Zr90cs} and \ref{fig:Zr92cs} for the $^{90}$Zr(p,$\gamma$)$^{91}$Nb and 
$^{92}$Zr(p,$\gamma$)$^{93}$Nb reactions, respectively. As mentioned in section \ref{sec:INTRO}, for the reaction $^{90}$Zr(p,$\gamma$)$^{91}$Nb 
there were two previous measurements that were not in agreement. The cross section measurement by Laird {\it et al.} \cite{Lai87} are presented in the main panel of 
Fig.~\ref{fig:Zr90cs} in open circles while the reaction rates presented by Roughton {\it et al.} \cite{Rou79} are shown in the inset of Fig.~\ref{fig:Zr90cs}.
In both Figs.~\ref{fig:Zr90cs} and \ref{fig:Zr92cs} the solid line presents theoretical calculations from the code NON-SMOKER \cite{Rau01}. These calculations represent the 
suggested reaction rates in the JINA REACLIB database \cite{REACLIB}. The dashed and dotted 
lines in both figures present theoretical calculations performed with the code TALYS \cite{Kon04}. The TALYS calculations were performed using default
parameters, except for the Optical Model Potential (OMP) and the Nuclear Level Densities (NLD). The semi microscopic OMP of Bauge, Delaroche, and Girod
\cite{Bau01} was used, which is a reparametrization of the standard microscopic OMP of Jeukenne, Lejeune, and Mahaux (JLM) \cite{JLM} to cover a
wider energy range of scattering data. The JLM potential was used in the NON-SMOKER calculations presented here. TALYS offers a variety of options for 
the description of the Nuclear Level Densities. In the present work, the use of the different NLDs gave similar results and for this reason only the lowest and highest cross section 
calculations are presented in the figures, as limits of these calculations. For both reactions the highest limit (dashed lines) was the result of using the NLD from the 
Generalized Superfluid Model (GSM) \cite{GSM1, GSM2}, while the lower limit (dotted lines) was calculated using the microscopic NLD from Hilaire's table (HT) \cite{HT}.

In both reactions, a good agreement between the results of the present work and the three calculations shown in Figs.~\ref{fig:Zr90cs} and \ref{fig:Zr92cs} is observed 
at the lowest energies, in the regions where the sensitivity is dominated by the proton width. The good agreement with the NON-SMOKER calculation remains also 
at higher energies, with significant deviations only present in the $^{92}$Zr(p,$\gamma$)$^{93}$Nb reaction above the neutron threshold. Larger discrepancies are observed when
comparing to the calculations using the TALYS code, although the calculation using the GSM nuclear level densities can reproduce the experimental data fairly well. 
Overall, NON-SMOKER yields a better description
of the data for both reactions. Since, it reproduces the
experimental data well within the uncertainties, the astrophysical
reaction rates obtained with NON-SMOKER are also confirmed. Therefore we
refrain from giving rate tables but just refer to \cite{Rau00,Rau01}. 

In comparing the results of the present work with previous experimental results, it is observed that the reaction rates obtained by Roughton {\it et al.} \cite{Rou79}
shown in the inset of Fig.~\ref{fig:Zr90cs} are in good agreement with the standard NON-SMOKER calculations which can describe the results of the present
work as well. A larger discrepancy is observed between the present results and the ones from Laird {\it et al.} \cite{Lai87}, especially at the higher energies. 
As mentioned in section \ref{sec:INTRO}, the latter authors used the method of $\gamma$ angular distribution measurements to extract the reaction cross section. 
This method relies on the detection of all individual $\gamma$ rays contributing to the cross section and presents the risk of missing low intensity $\gamma$ rays 
that are below the detection (and background) limits of the experimental setup.
With the technique used in the present work we were able to extract average $\gamma$ multiplicities, $<$M$>$, as described in \cite{Sim13}. For the $^{90}$Zr(p,$\gamma$)$^{91}$Nb reaction, 
the average multiplicity increases with beam energy getting to a value of approximately $<$M$>$=~3 for proton beam energy of 5~MeV. This result supports the
observation that the $\gamma$-angular distribution data of Laird {\it et al.} are in good agreement with our measurement at the low beam energies and the deviation
increases as the beam energy and average multiplicity are increased.

The astrophysical calculations performed in the present work, as outlined in Sec.~\ref{sec:INTRO}, did not show any significant contribution of the studied reactions to the production of the 
p nuclei $^{92, 94}$Mo. Since the results confirmed the NON-SMOKER calculations that were already used in the astrophysical model, no change was observed
in the final abundances and the puzzle of the underproduction of the neutron deficient molybdenum and ruthenium isotopes remains.


\section{Conclusions}
\label{sec:CONC}

The present work reports on two cross section measurements of the reactions $^{90}$Zr(p,$\gamma$)$^{91}$Nb and 
$^{92}$Zr(p,$\gamma$)$^{93}$Nb. The measurements were performed using the NSCL SuN detector at the University of 
Notre Dame in the energy range between 2.0 and 5.0~MeV. This energy range covers the majority of the Gamow window for
the two reactions. The results were found to be in good agreement with one of the previous measurements for the
$^{90}$Zr(p,$\gamma$)$^{91}$Nb reaction. The results were also in excellent agreement with the predictions of the 
code NON-SMOKER, the calculations of which are commonly used in astrophysical models. Somewhat larger 
discrepancies were observed when using the code TALYS for the cross section calculations. 
This work provides a confirmation of the reaction rates provided by the code NON-SMOKER and used in the JINA 
REACLIB database for the two reactions studied. However, no generalization for other reactions can be inferred from the present work.

The authors would like to thank Hendrik Schatz from
Michigan State University for providing the input distributions
used in Ref.~\cite{Rap06}.
The authors also gratefully acknowledge the support of NSL operations staff at the University
of Notre Dame during the two experiments.
This work was supported by the National Science Foundation under Grant No. PHY-1102511 and 
PHY 08- 22648 (Joint Institute for Nuclear Astrophysics). TR acknowledges the support of 
the Swiss NSF, the EUROCORES EuroGENESIS research program, the ENSAR/THEXO European FP7 program
and the Hungarian Academy of Sciences.



\end{document}